\documentclass[10pt, conference, letterpaper]{IEEEtran}
\usepackage{graphicx}
\usepackage{caption}
\usepackage{xspace}
\usepackage{amsmath}
\usepackage{amsthm}
\usepackage{amssymb}
\usepackage{xfrac}
\usepackage{subfigure}
\usepackage{balance}
\usepackage{algorithm}
\usepackage[noend]{algpseudocode}

\newcommand{\imin}{$I_{min}$\xspace}
\newcommand{\imax}{$I_{max}$\xspace}
\newcommand{\I}{$I$\xspace}

\newcommand{\etal}{\emph{et~al.}\xspace}
\begin{document}

\title{
Multiple Redundancy Constants with Trickle
}
\author{
	\IEEEauthorblockN{
		Titouan Coladon\IEEEauthorrefmark{1},
		Mali\v{s}a Vu\v{c}ini\'{c}\IEEEauthorrefmark{1}\IEEEauthorrefmark{5}\IEEEauthorrefmark{3},
		Bernard Tourancheau\IEEEauthorrefmark{1}
	}
	
	\IEEEauthorblockA{ 
		\IEEEauthorrefmark{1}Grenoble Alps University, CNRS Grenoble Informatics Laboratory UMR 5217, France.
	}
	
	\IEEEauthorblockA{
		\IEEEauthorrefmark{5}STMicroelectronics, Crolles, France.\\
			}
	    \IEEEauthorblockA{
        \IEEEauthorrefmark{3}BSAC, University of California, Berkeley, CA, USA.
        }
Email: \{firstname.lastname\}@imag.fr.
}

\hyphenation{sen-sors net-works low-power manu-facturers lite-rature}

\maketitle

\begin{abstract} 
Wireless sensor network protocols very often use the Trickle algorithm to govern 
information dissemination. For example, the widely used IPv6 Routing Protocol for Low-Power and 
Lossy Networks (RPL) uses Trickle to emit control packets. We derive an analytical model of Trickle to
take into account multiple redundancy constants and the common lack of synchronization among nodes.
Moreover, we demonstrate message count
unfairness when Trickle uses a unique global redundancy constant because nodes with less neighbors
transmit more often. Consequently, we propose a heuristic algorithm that calculates 
a redundancy constant for each node as a function of its number of neighbors. 
Our calculated redundancy constants reduce unfairness among nodes by distributing more 
equally the number of transmitted messages in the network. 
Our analytical model is validated by emulations of constrained devices running the Contiki Operating System 
and its IPv6 networking stack.
Furthermore, results very well corroborate the heuristic algorithm improvements.
\end{abstract}

\noindent

\vspace{-0.25cm}

\section{Introduction}
\label{intro}
The Trickle algorithm is a timer based control algorithm relying on recursive doubling time intervals
and ``polite gossip'' policy. It quickly propagates updates in the network but avoids unnecessary
transmissions. It was initially proposed for system code versioning \cite{trickle-paper} in Wireless 
Sensor Networks (WSN). 
Due to its wide-spread use,
it has been separately standardized in RFC~6206 \cite{trickle-rfc}. 
Most notably, the Routing Protocol 
for Low Power and Lossy Networks (RPL)~\cite{rpl} utilizes Trickle for topology maintenance.
Also, Multicast Protocol for Low Power and Lossy Networks 
\cite{multicast} and other protocols \cite{melete, ctp, deluge} build upon it, 
leveraging Trickle's benefits. 
This makes the understanding of its behavior
crucial for performance optimization of control overhead. 


In this paper, we model and study the operation of Trickle, which leads to a better understanding
of the impact of its redundancy constant. We demonstrate that the usage of a unique redundancy constant for the whole
network leads to communication unfairness when the underlying topology density is not homogeneous. The root cause of this unfairness
is the increased probability of transmission of nodes with less neighbors in their radio vicinity. This results in uneven transmission
load, e.g. message count, across the network. Moreover, in battery powered networks, these nodes with a higher transmission probability will cease functioning  sooner because of on board energy depletion as broadcasting is very expensive in WSNs.
We model the individual transmission probabilities with individual node redundancy constants across the network.  
The model with multiple redundancy constants can be numerically resolved for arbitrary topologies but also simplified to closed-form in
specific cases, outside of the scope of this paper.

From the model's results, we propose a simple heuristic algorithm in order to improve Trickle fairness by a local computation of each
redundancy constant as a function of the number of neighbors. We demonstrate the resulting improvements in terms of
transmission load balance  both by leveraging our analytical model results and by emulating constrained-node networks running the full Contiki Operating System network stack. 

The main contributions of the paper are the following:
\begin{itemize}
\item A new probabilistic model estimating the message count and average transmission probabilities of individual nodes in steady state networks. This model works for
arbitrary topologies without any synchronization requirement, and accounts for multiple redundancy constants among nodes,
\item A demonstration of transmission load unfairness in networks utilizing a fixed redundancy constant among nodes,
\item A new algorithm improving fairness in the network by locally computing the redundancy constants as a function of the number of neighbors in the node's radio vicinity,
\item A validation of the model and an evaluation of the proposed algorithm improvements. The emulation uses highly accurate instruction-level execution of the binary file that contains our code and the Contiki network stack and normally runs on real hardware.
\end{itemize}

The paper is structured as follows. Section \ref{trickle} reviews the Trickle algorithm. In Section \ref{related-work}, we discuss the 
related work in  details. Section \ref{model} presents our probabilistic model design. We validate the model and
discuss unique redundancy constant unfairness  in Section \ref{validation}. In Section \ref{computation-k}, we present our heuristic algorithm
that locally computes each redundancy constant and discuss its achieved improvements. Finally, we conclude and 
discuss future work in Section \ref{conclusion}.

\section{The Trickle Algorithm}
\label{trickle}
The main idea of the Trickle algorithm is on one hand to exponentially reduce the amount of control traffic
in the network, while there are no detected inconsistencies in a given state. On the other hand, once an inconsistency has been detected it quickly propagates the new information state. 
Naturally, the "consistency notion" 
is defined by the protocol or application actually using Trickle. For instance, in the case of IETF RPL routing
protocol, consistency is checked by comparing the advertised DODAG state in the network to the local
one. In the case of firmware updates a similar verification is usually performed between  software
versions.

Trickle splits time into intervals of variable length where transmissions may occur following Trickle's rules. The three parameters to configure Trickle are:
i) \imin, the minimum interval size; ii) \imax, the maximum interval size expressed as the number of
times the minimum interval may double; iii)~$K$, the redundancy constant.
  
\begin{figure}[tb]
    \includegraphics[width=1\columnwidth]{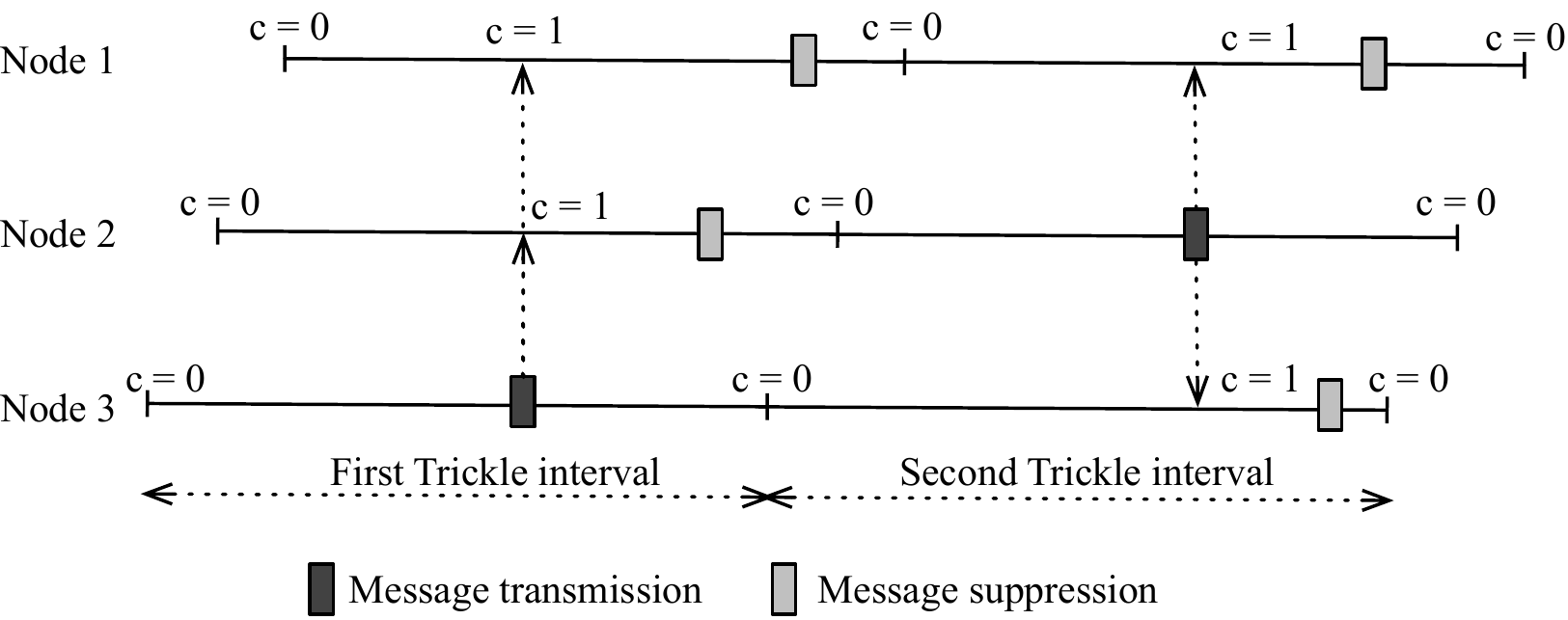}
    \caption{Example of Trickle algorithm in steady state with the lack of synchronization among nodes. Redundancy constant $K=1$, and 
     all nodes are neighbors.}
    \label{trickle_example}
    \vspace{-0.5cm}
\end{figure}

A node following the Trickle algorithm increments a local counter $c$ for each consistent reception.
The node transmits at instant $t$ if:
\begin{equation}
c < K,
\end{equation}
that is, if the number of consistent receptions is smaller than the redundancy constant. Counter $c$ is reset to
zero at the beginning of each interval. Instant~$t$ at which Trickle decides if it is 
going to transmit is selected randomly from the uniform interval [$\frac{1}{2}$\I,~\I), where $I\in~\{I_{min}~\times~2^{n}~\mid
n \in \mathbb{N}_{0}, n \le I_{max}\}$. Interval~\I is doubled upon its
expiration by incrementing $n$. When a node detects inconsistency, $n$ becomes $0$, which sets
interval \I to~\imin. In this paper, though, we model Trickle networks in steady state, such that
\I~=~\imin~$2^{I_{max}}$, and focus on the effect of the redundancy constant. Fig. \ref{trickle_example} 
illustrates an example Trickle operation in steady state and $K=1$. As soon as $c \ge K$, transmissions are suppressed.
Note that the Trickle intervals among nodes are not necessarily synchronized. 

\section{Related Work}
\label{related-work}
Due to its wide-spread use, the Trickle algorithm has been subject to many studies 
\cite{becker, kermajani, trickle-f, meyfroyt, rpl-topologyconstruction, trickle-l2, kermajani2014network}.
Becker~\etal \cite{becker} develop a model to study the propagation time of new information in a network using
Laplace transforms. 

Meyfroyt~\etal \cite {meyfroyt} recently published a model generalizing the algorithm by introducing the \emph{listen-only}
parameter $\eta$. In the standardized version of Trickle \cite{trickle-rfc} and the original paper \cite{trickle-paper} $\eta=\frac{1}{2}$
and is introduced in order to avoid broadcast storms in unsynchronized networks at the beginning of intervals, by
forcing nodes to keep listening before attempting transmissions (i.e. listen-only period).
The authors demonstrate that using a short listen-only period provides advantage in terms of smaller propagation
time, but in the same time increases the number of transmitted messages in the network. They derive the cumulative
distribution function of inter-transmission times for large number of nodes in a steady state, unsynchronized, single cell network.

Kermajani \etal \cite{kermajani} approach the problem of estimating the Trickle message count in steady state
  by deriving the average probability $P$ that a node in the network will transmit in a given interval. Then, the
average message count in a given interval is simply $N \times P$, where N denotes the number of nodes in the network.
In respect to the model of Meyfroyt \etal \cite{meyfroyt}, the approach implicitly supports multi-cell topologies.

A common point on published Trickle models \cite{becker, meyfroyt, kermajani} (and deployments)
is that they all consider a
unique, fixed redundancy constant among nodes. The unique redundancy constant leads to unfairness as nodes with less 
neighbors have less incoming packets and thus a higher probability to transmit, and will therefore deplete their available energy source sooner.
Note that Trickle messages translate to L2 broadcast frames that are very expensive in WSNs.

Other authors \cite{trickle-f, rpl-topologyconstruction, trickle-l2, kermajani2014network} have studied Trickle and its performance in the 
specific use case of RPL and how it affects the convergence and route optimality of the DODAG building process.

\section{Modelling the Trickle Message Count}
\label{model}
Our model calculates the message count of individual nodes in a steady state Trickle-based network.
Similarly to Kermajani~\etal \cite{kermajani}, we derive the average probability of transmission. However, we use a different decomposition that allows us to extend their approach
in order to calculate per node probabilities, rather than the network average. Hence we can give insights on the fairness of the algorithm.
To render the model more practical, we do not make explicit assumptions on topology. Rather, we assume that each node $i$ is
able to 
know 
its number of neighbors $y_i$. Therefore, our model requires  for each node its redundancy constant $K_i$, 
and its neighbors list. 

The main idea of our analysis is to express the average probability of transmission of a node as a function of the transmission probabilities of its
neighbors. This will yield a system of $N$ equations with $N$ unknowns, where $N$ is the number of nodes in the network, that
can be numerically resolved. 

\textbf{Distribution of Transmission Times.}
In networks with synchronized Trickle intervals, transmission times simply follow a uniform distribution \cite{kermajani}. 
However, with the lack of synchronization, this does not longer hold. 
Let $X_1,  \ldots ,X_{y_i}$ be i.i.d. random variables of uniform distribution modeling the transmission time positions of the $y_i$ nodes into an interval of length $I$.
Let $T$ be the selected transmission time of node $i$, $T \in [\frac{1}{2}\I,~\I]$.
Let $Y_T$ denote the number of selected transmission times before $T$.
Let $n$ be the positive integer that denotes the position of transmission time of node $i$ in the set of increasingly ordered transmission times.
The probability that $n$ is selected by node $i$ and by its neighbors
is equal to $P(Y_T=n-1)$.
$Y_T$~can be shown to follow a binomial distribution with parameters $y_i$ and $\displaystyle\frac{T}{I}$.

\subsection{Probabilistic Model}
The average probability that node $i$ will send a message in a given interval is denoted $P_{TX}[i]$.
A node will surely transmit in the case where its number of neighbors $y_i$ is less than its redundancy constant $K_i$, because the counter $c$ can never
reach $K_i$. Otherwise, the node transmits in two cases: i) if it selects any of the first $K_i$ transmission times, ii) if it selects any of the last
$y_i+1-K_i$ transmission times and at most $K_i - 1$ of its neighbors have already transmitted. 
Consequently, $P_{TX}[i]$, can be written as follows:
\begin{equation*}
    P_{TX}[i]= \begin{cases}
    1, \hfill y_i < K_i \\
    P_F[i] + P_{LO}[i], \quad  y_i \geq K_i.
    \end{cases}
\end{equation*}
where:
\begin{itemize}
 \item $P_F[i]$ is the probability that node $i$ selects one of the first $K_i$ transmission times. We can find
 this probability simply as: $P_F[i] = \displaystyle\sum \limits_{n=0}^{K_i-1}P(Y_T=n)$.
 \item $P_{LO}[i]$ is the probability that node $i$ selects any of the last $y_i+1-K_i$ transmission times and at most $K_i-1$ nodes, with a lower transmission time than node $i$, will transmit before it. We refer to this probability as the last opportunity transmission probability.
 This probability depends on $P_{TX}[j]$, where $j$ is a neighbor of node $i$.

\end{itemize}




\textbf{Last Opportunity Transmission Probability.}
We are considering the case where node $i$ selected one of the last $y_i+1-K_i$ transmission times.
The probability that at most $K_i-1$ nodes, with a lower transmission time than node $i$ transmit before, depends
on transmission time of node~$i$. 
We will compute the probability $P_{LO}[i]$ by conditioning on transmission time of node $i$.
As $Y_T \in \{K_i,  \ldots , y_i\}$, $P_{LO}[i]$ can be derived as:
 \begin{equation}
P_{LO}[i]=\displaystyle\sum \limits_{n=K_i}^{y_i} P_{LO}[ i \ | \ Y_T= n]  \times P(Y_T=n). 
 \end{equation}
 
Let $B_{set}$ be the set of $n$ neighbors of node $i$, denoted by \{$1$,  \ldots , $n$\} whose transmission times are lower than the one of node $i$.
Let $\Re$ be the set composed of $\dbinom{y_i}{n}$ possible sets of nodes $B$ that possibly match $B_{set}$. \newline
Therefore, $P_{LO}[i \ | \ Y_T= n]$ can be obtained as:
\begin{equation}
 P_{LO}[i | Y_T= n]=\displaystyle\frac{1}{\dbinom{y_i}{n}}\displaystyle\sum_{\substack{B \in \Re}} P_{LO}[i | Y_T= n \land B_{set}=B]
\end{equation}
The probability that node $i$ transmits in this case, $P_{LO}[i \ | \ Y_T~=~n \ \land \ B_{set}=B]$, is the probability that at most $K_i-1$ nodes of $B$ transmit before:
\begin{equation}
 P_{LO}[i \ | \ Y_T= n \ \land \ B_{set}=B ]=\displaystyle\sum \limits_{j=0}^{K_i-1}\gamma(j,n,B),
\end{equation}
where $\gamma(j,n,B)$ denotes the probability that $j$ nodes of the set $B=\{1,  \ldots , n\}$ transmit before node $i$. 
By definition of $P_{TX}$, we have:
\begin{equation*}
\gamma(0,n,B)=\displaystyle\prod \limits_{l=1}^{n}(1-P_{TX}[l]).
\end{equation*}
More generally:
\begin{multline*} 
\gamma(k,n,B) = \displaystyle\sum \limits_{i_1=1}^{n}\displaystyle\sum_{\substack{i_2=1 \\ i_2 \neq i_1}}^{n}  \ldots  \displaystyle\sum_{\substack{i_k=1, \\ i_k \neq i_1, \ldots , i_{k-1}}}^{n} [P_{TX}[i_1] \times  \\ \ldots  \times P_{TX}[i_k] \times \displaystyle\prod_{\substack{l=1, \\ l\neq i_1,  \ldots , i_k}}^{n}(1-P_{TX}[l])].\newline
\end{multline*}
By re-organizing the sums, this leads to:
\begin{equation}
 P_{LO}[i]=\displaystyle\sum \limits_{n=K_i}^{y_i} P(Y_T=n) \times \displaystyle\frac{1}{\dbinom{y_i}{n}}\displaystyle\sum_{\substack{B \in \Re}} \displaystyle\sum \limits_{j=0}^{K_i-1}\gamma(j,n,B).
\end{equation}

For example, let us consider the case where
node $i$ has $y_i=4$ neighbors, $\{a,b,c,d\}$, and $n=2$ of them have selected
a transmission time lower than its own. We can have $\dbinom{4}{2}$ different sets of neighbors $B_1=\{a,b\},B_2=\{a,c\},B_3=\{a,d\},
B_4=\{b,c\},B_5=\{b,d\},B_6=\{c,d\}$,
whose nodes have a lower transmission time.
For instance, when $B=B_1=\{a,b\}$ and  $K_i=2$, 
node $i$ will transmit either if $j=0$ nodes of $B$ transmit, or if $j=1=K_i-1$ nodes of $B$ transmit.
The probability that $j$ nodes of the set $B$ transmit is $\gamma(j,n,B)$.

We now have $P_F[i]$ and $P_{LO}[i]$ expressed according to $\{P_{TX}[j] \ | \ \text{$j$ is a neighbor of node $i$} \}$.
In order to find $P_{TX}[i] \ \forall i \in \ \{1,  \ldots , N\}$, we need to solve the $N$ equations with $N$ unknowns, i.e. the system $P_{TX}[i]=P_F[i]+P_{LO}[i]$. The solutions of the system are the average probabilities of transmission for each node in the network
and in the same time the average message count per node during one interval. In its general form, the system of equations models 
arbitrary network topologies and can be resolved numerically. For specific topologies which are outside of the scope of this paper,
a closed form solution may be obtained.

Due to its complexity, the general form of the model does not allow a direct practical implementation for constrained devices. 
However, its numerical resolution gives precise insights on node behavior and where the imbalance in the network occurs. 
Based on this, in Section \ref{computation-k} we propose a practical, heuristic approach that is easily computable locally.

\section{Model Validation and Trickle Unfairness}
\label{validation}

To validate our model we implemented a tool resolving the model in Python and Sage \footnote{http://sagemath.org}, an open-source computational software program.
We emulate the Tmote Sky sensor motes running Contiki Operating System, by using the MSPSim emulator, and the Cooja simulator,
in order to obtain real world results of Trickle. We use the Trickle application level library code available in Contiki.
The same binary file used for emulations runs on real hardware without any modifications.
 Note that due to
the use of the emulator, some imperfections
of results in respect to the real deployments come from the Unit Disk Graph (UDG) radio model in Cooja. We validate the model
for 49 node networks: i)~using a 7 $\times$ 7 grid topology, to demonstrate unfairness, ii)~using a randomly generated topology, to demonstrate the validity of the model for the general case. Transmission range for the grid topology was $R=\sqrt{2}$, with the average node degree of 6.37.
In the case of randomly generated topology, the average node degree was 3.92.
We average emulation results over 30 runs, and calculate
the model for the same topology based on the list of neighbors of each node. We count the number of transmissions of each node over
10 steady state Trickle intervals of 16 seconds. We calculated 95 \% confidence intervals but do not
present them on the graphs for the sake of clarity, as they are graphically indistinguishable from
the plotted averages.

\begin{figure}[tb]
	\centering
		\includegraphics[width=0.7\columnwidth]{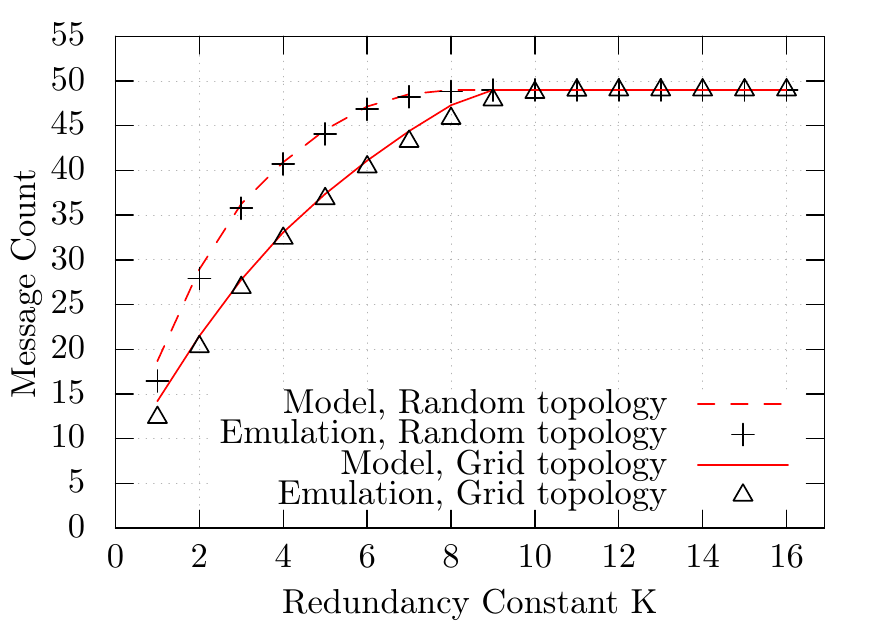}
		\label{const_k_grid}
		\label{const_k_uniform}
\caption{Message count in networks with fixed redundancy constant.
}
\label{validation_const_k}
\vspace{-0.5cm}
\end{figure}

Fig. \ref{validation_const_k} presents the results for the total message count in the network using a unique, fixed redundancy constant
among nodes. The model accurately predicts the number of messages in the network. 
Numerical values of maximum, minimum probabilities, variance and their comparison with the emulation results are shown in
Tables~\ref{table_1_6} and \ref{table_1_6_uniform}.
The imperfections of the model come from the fact that with the lack of synchronization of Trickle intervals among nodes, if we consider a node with $y$ neighbors, 
$y$ is only the mean number of transmission times that can occur during one interval.
Also, the assumption that transmission probabilities of nodes are independent events does not hold. As discussed by
Kermajani \etal \cite{kermajani}, a transmission performed by a node causes the increment of the counter $c$ and therefore 
decreases the transmission probability of its neighbors. Nevertheless, emulation results are obtained by running the binary file that normally executes on real hardware. They show that our model 
provides accurate estimations of the average transmission probabilities of nodes in the network, and consequently of the message count.

\begin{table*}[tb]
 \caption{Model and emulation results on $7\times 7$ grid, $R=\sqrt{2}$, for $K~\in \{1,2,3,4,5,6\}$} 
 \begin{tabular}{|p{1.8cm}||p{0.82cm}|p{0.82cm}||p{0.82cm}|p{0.82cm}||p{0.82cm}|p{0.82cm}||p{0.82cm}|p{0.82cm}||p{0.82cm}|p{0.82cm}||p{0.82cm}|p{0.82cm}|}
   \hline  
   results / redundancy constant & model $K=1$  & emul. $K=1$  & model  $K=2$  & emul.  $K=2$ & model $K=3$  & emul. $K=3$  & model $K=4$  & emul.  $K=4$  & model $K=5$  & emul.  $K=5$  & model  $K=6$ & emul.  $K=6$ \\
   \hline  \hline
   max probability & 0.673 & 0.606 & 0.887 & 0.896 & 0.980 & 0.983 & 0.999 &  1.0 & 0.999 &  1.0  & 0.999 & 1.0 \\
   \hline
   min probability & 0.070 & 0.05 & 0.084 & 0.05 & 0.116 & 0.153 & 0.173 & 0.22  & 0.295 & 0.38 & 0.501 &  0.493 \\
   \hline
   variance & 0.03217 & 0.02466 & 0.06402 & 0.05030 & 0.08261 & 0.05736 & 0.08553 & 0.06077 & 0.06401 & 0.05158 & 0.03268 &  0.03339 \\
   \hline
 \end{tabular}
 \label{table_1_6}
 \vspace{-0.2cm}
\end{table*}

\begin{table*}[tb]
 \caption{Model and emulation results on 49 node random topology and 3.92 average node degree, for $K~\in \{1,2,3,4,5,6\}$} 
 \begin{tabular}{|p{1.8cm}||p{0.82cm}|p{0.82cm}||p{0.82cm}|p{0.82cm}||p{0.82cm}|p{0.82cm}||p{0.82cm}|p{0.82cm}||p{0.82cm}|p{0.82cm}||p{0.82cm}|p{0.82cm}|}
   \hline  
   results / redundancy constant & model $K=1$  & emul. $K=1$  & model  $K=2$  & emul.  $K=2$ & model $K=3$  & emul. $K=3$  & model $K=4$  & emul.  $K=4$  & model $K=5$  & emul.  $K=5$  & model  $K=6$ & emul.  $K=6$ \\
   \hline  \hline
   max probability & 0.853 & 0.773 & 0.999 & 0.996 & 0.999 & 1.0 & 0.999 & 1.0 & 0.999 & 1.0 & 0.999 & 1.0 \\
   \hline
   min probability & 0.070 & 0.056 & 0.033  & 0.083 & 0.095 & 0.146 & 0.233 & 0.28 & 0.427 & 0.396 & 0.663 & 0.623 \\
   \hline
   variance & 0.04783 & 0.03346 & 0.08453 & 0.07687 & 0.08518 & 0.07773 & 0.05276 & 0.04963 & 0.0237 & 0.02392 & 0.00772 & 0.00750 \\
   \hline
 
 \end{tabular}
 \label{table_1_6_uniform}
 \vspace{-0.3cm}
\end{table*}

\begin{figure}[tb]
	\centering
	\subfigure[][K~=~1]{
		\includegraphics[width=0.4\columnwidth]{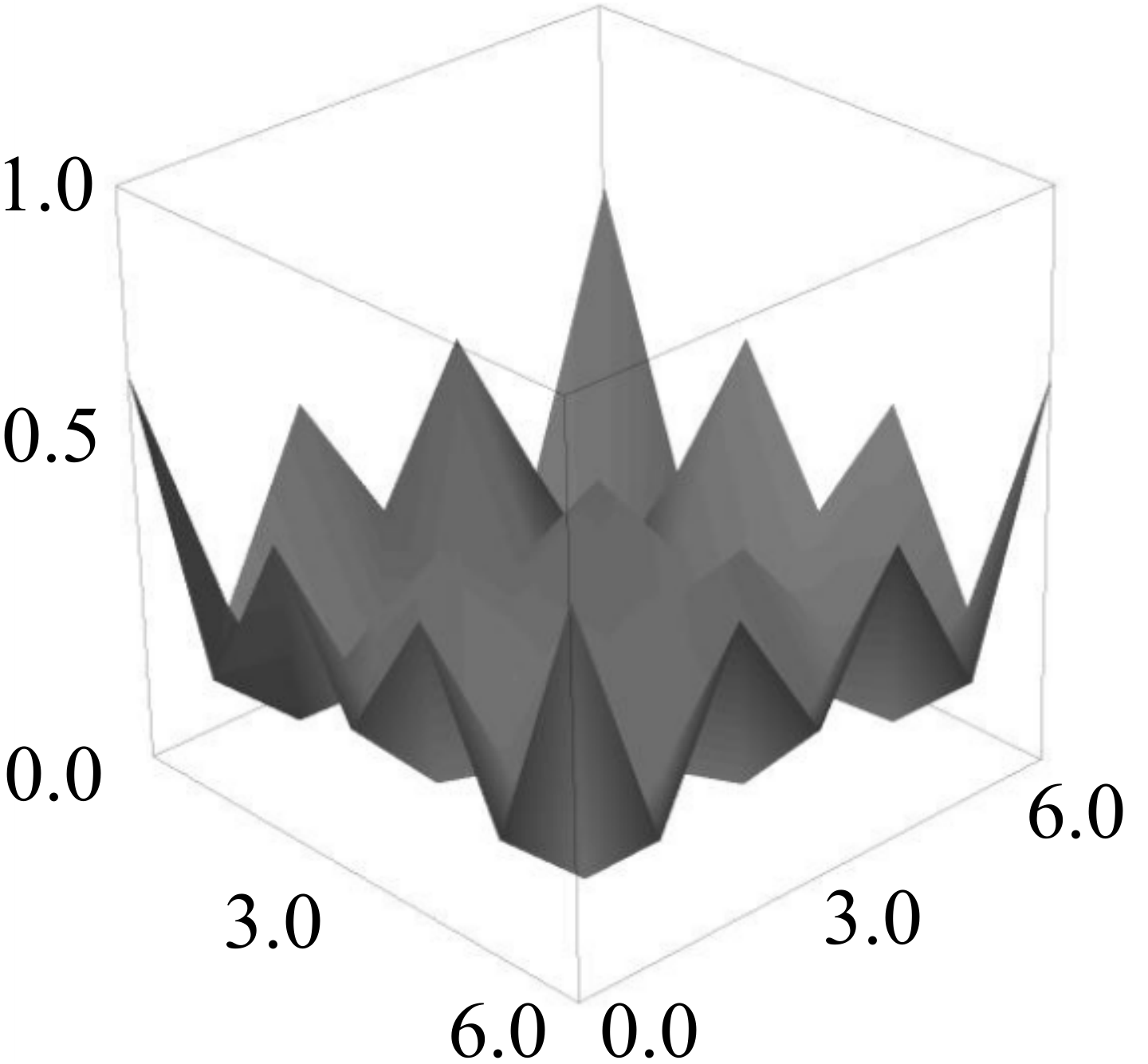}
		\label{3d_k1}
	}
	\subfigure[][K~=~2]{
		\includegraphics[width=0.4\columnwidth]{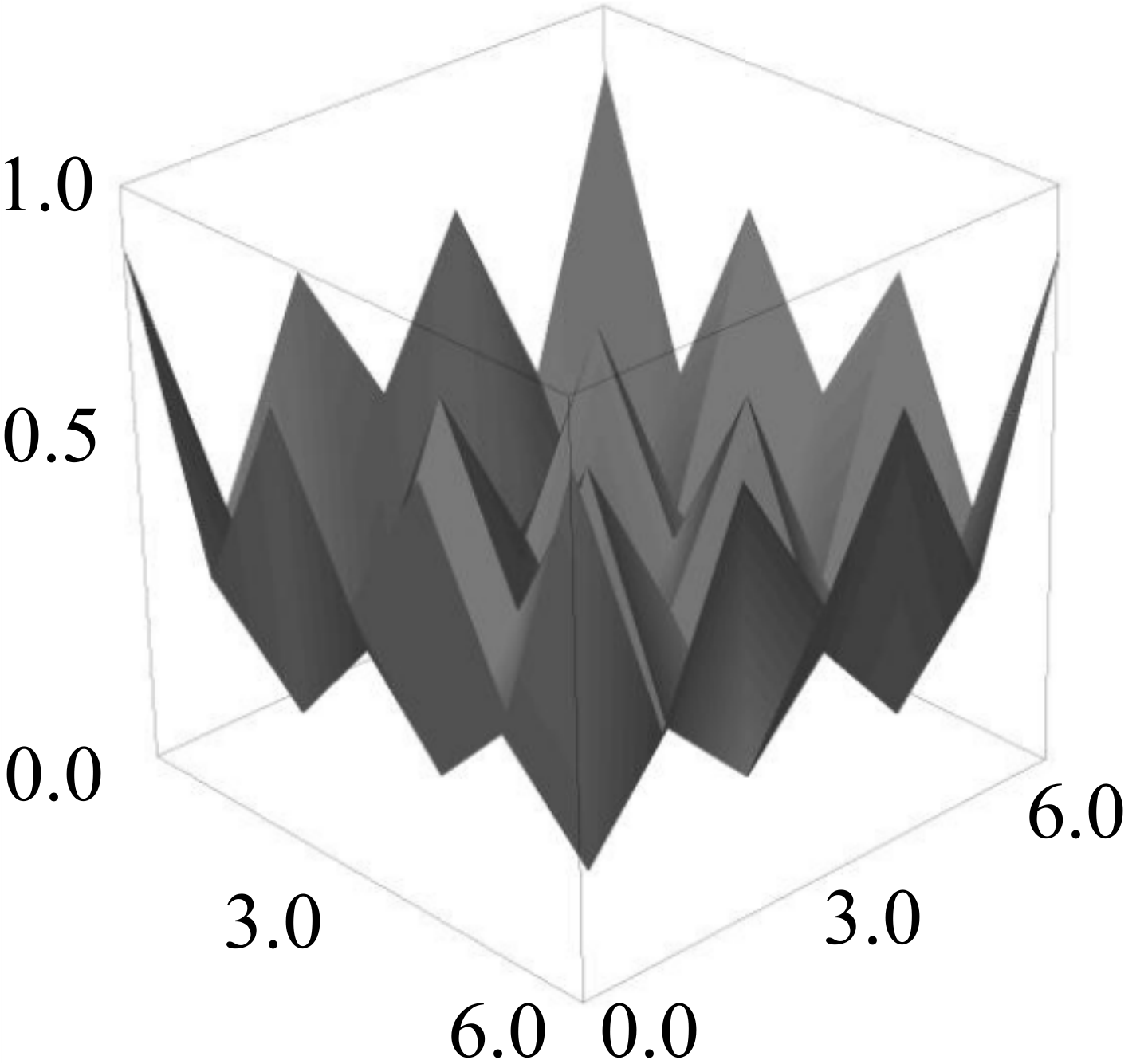}
		\label{3d_k2}
	}
	\subfigure[][K~=~3]{
		\includegraphics[width=0.4\columnwidth]{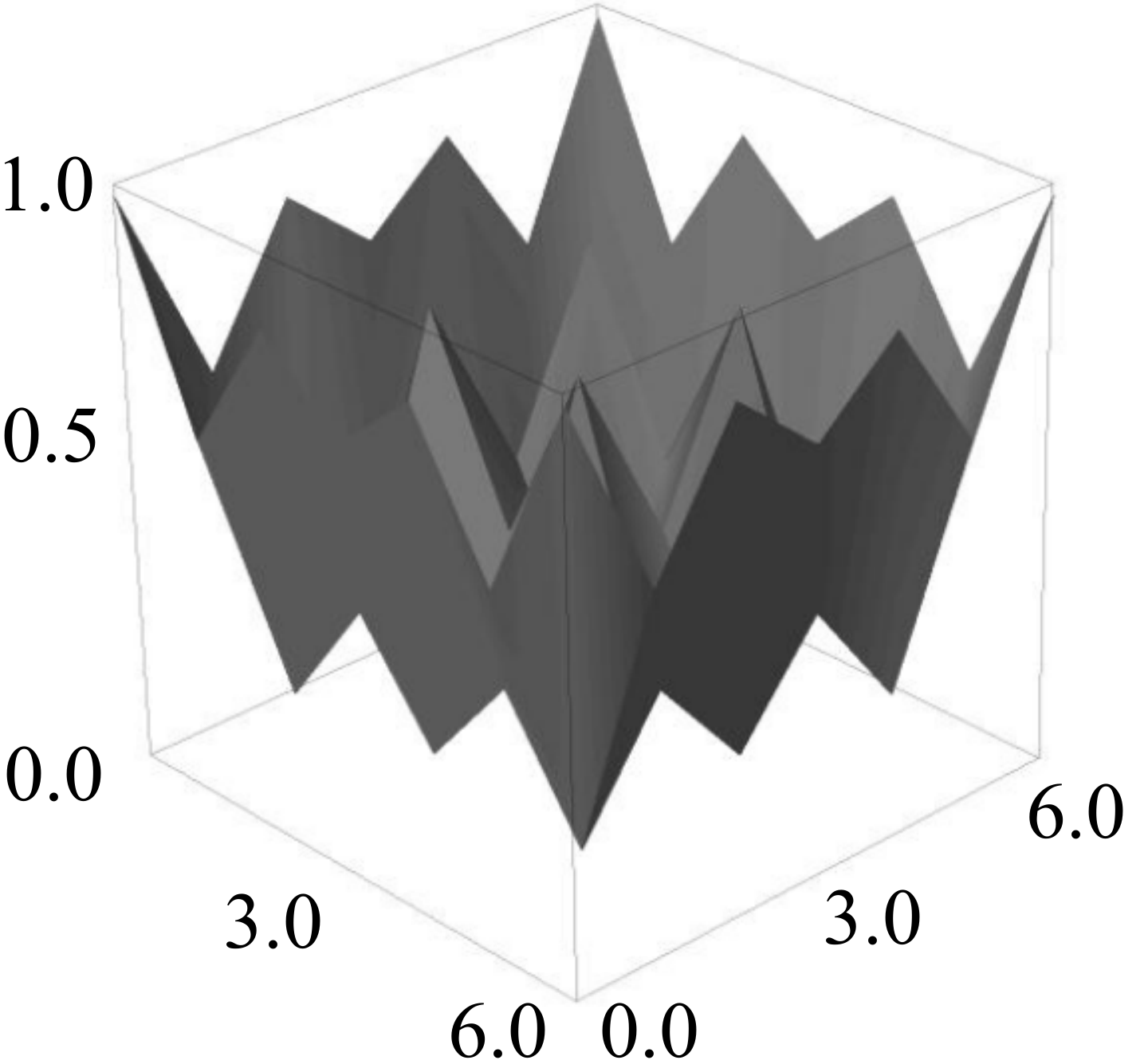}
		\label{3d_k3}
	}
	\subfigure[][K~=~4]{
		\includegraphics[width=0.4\columnwidth]{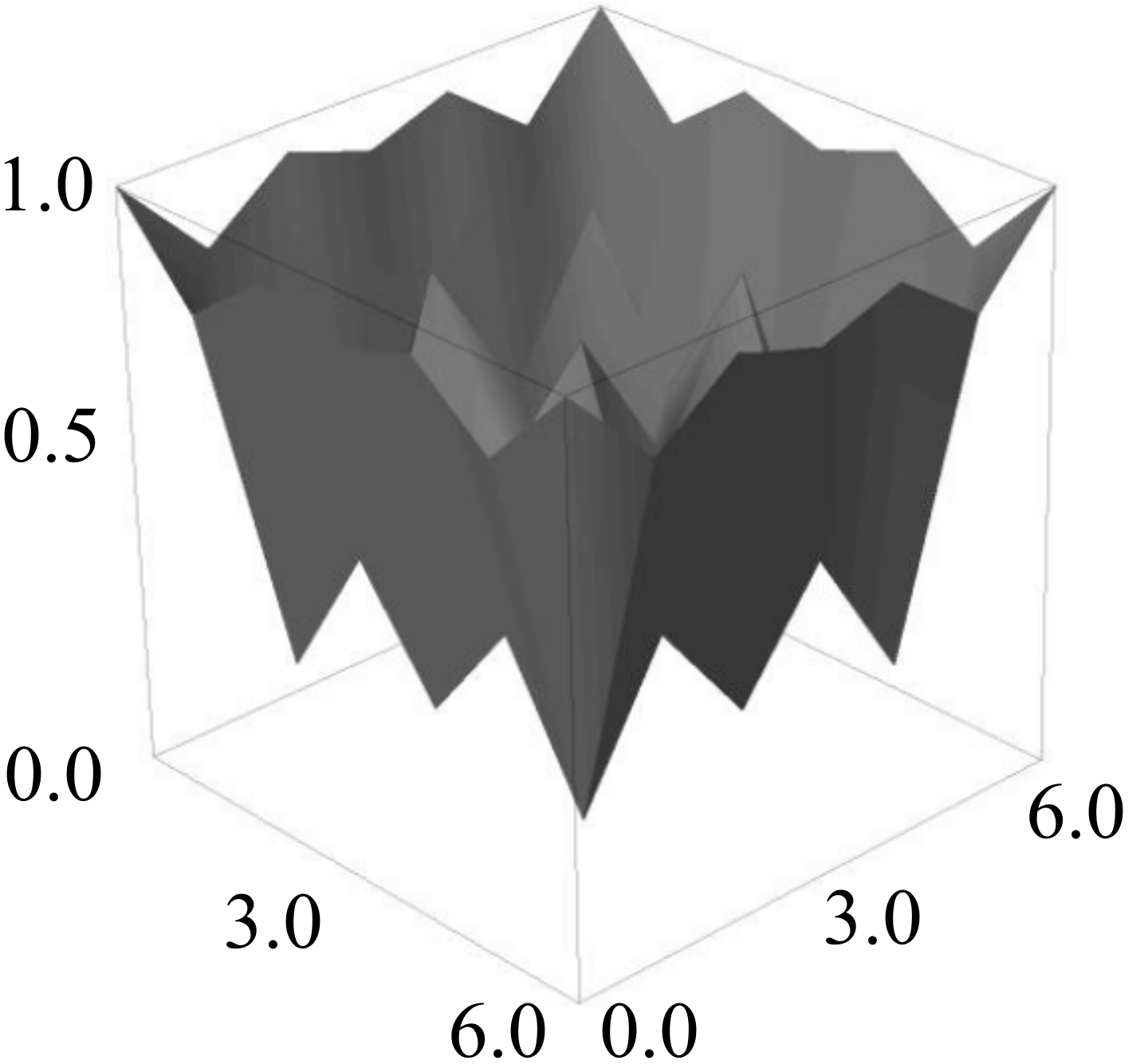}
		\label{3d_k4}
	}
\caption{Transmission probability of nodes in the grid estimated by the model for fixed redundancy constant.
}
\label{3d_const_k}
\vspace{-0.5cm}
\end{figure}

\textbf{Trickle Unfairness with Fixed Redundancy Constant.}
\label{unfairness}
As most real world deployments using Trickle utilize a unique, fixed redundancy constant among nodes,
we demonstrate the transmission unfairness that arises due to the heterogeneous network 
topologies, as nodes do not have the same number of neighbors. Fig. \ref{3d_const_k} presents the three 
dimensional graphs on probability of transmission calculated by our model for the grid topology, where the
effects are easily noted. We present results for $K \in \{1,2,3,4\}$, as the probabilities quickly approach
1.0 for larger $K$ (average node degree of 6.37). Inside the grid with 8 neighbors ($R=\sqrt{2}$), for $K=1$, 
we can see that the nodes have average transmission probabilities
of approximately 0.2, while the nodes on the edges with 5 neighbors have around 0.5, and the nodes in the corners with 3 neighbors on the
average transmit with the probability of approximately 0.7. 
Increasing the redundancy constant increases the transmission probabilities in the network. However, as the number of neighbors can
be considered fixed, nodes with less neighbors are affected more and their transmission probabilities increase faster in respect to those in the middle 
of the grid. This can be best seen for the case $K=4$ in Fig. \ref{3d_const_k}\subref{3d_k4}, as the nodes in the corners of the grid transmit with probability
1.0, while nodes on the edges have probability 0.85, and nodes inside the grid have probability~0.45.

\begin{figure*}[tb]
	\centering
	\subfigure[][Grid, K~=~2]{
		\includegraphics[width=0.7\columnwidth]{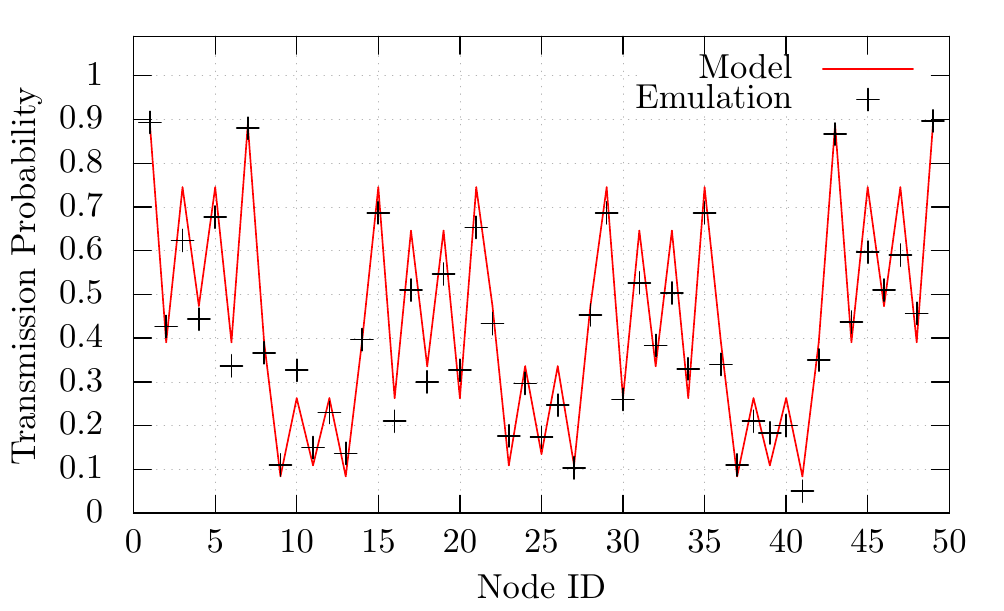}
		\label{per_node_k2_grid}
	}
	\subfigure[][Grid, K~=~3]{
		\includegraphics[width=0.7\columnwidth]{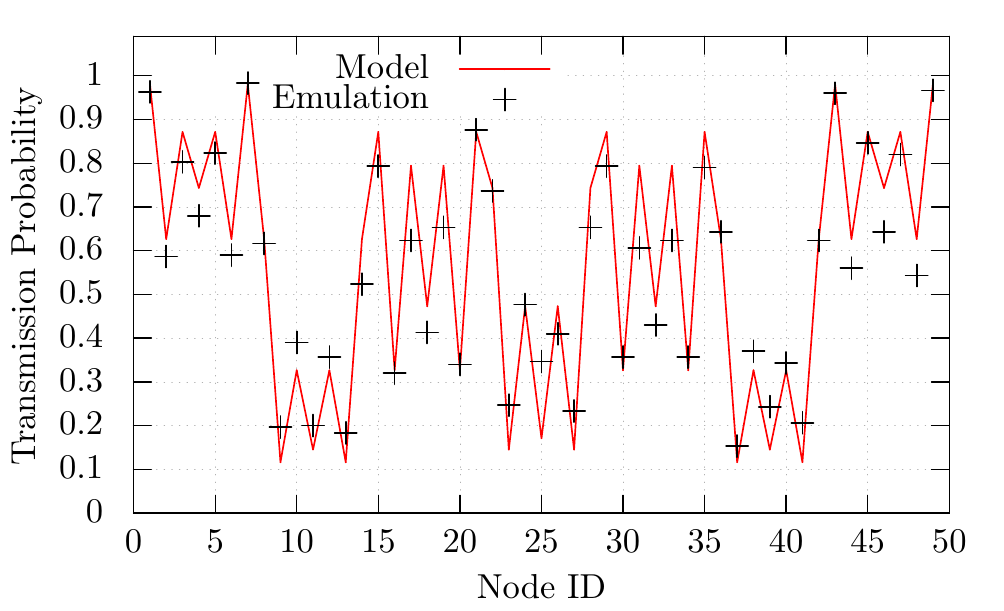}
		\label{per_node_k3_grid}
	}\\
	\subfigure[][Random, K~=~2]{
		\includegraphics[width=0.7\columnwidth]{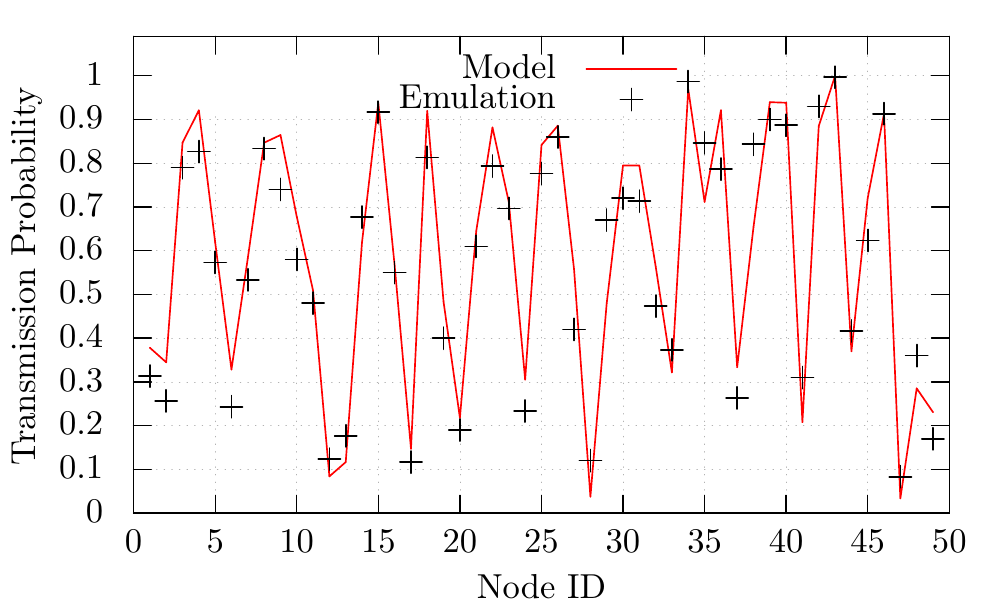}
		\label{per_node_k2_uniform}
	}
	\subfigure[][Random, K~=~3]{
		\includegraphics[width=0.7\columnwidth]{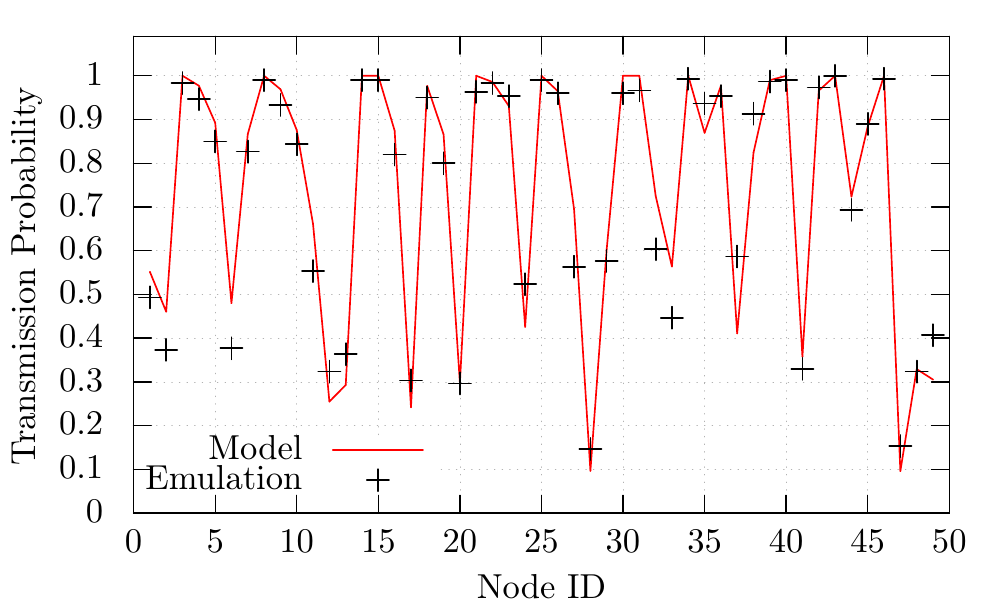}
		\label{per_node_k3_uniform}
	}
\caption{Transmission probability of nodes in the grid and random topology.
}
\label{per_node_grid_uniform}
\vspace{-0.5cm}
\end{figure*}

To validate the estimations of our model, we have confronted the results with emulations.
Fig. \ref{per_node_grid_uniform} presents the comparison for $K~\in~\{2, 3\}$, in the grid and the randomly generated topology. 

\section{Local Computation of the Redundancy Constant to Improve Fairness}
\label{computation-k}
As discussed, the average transmission probability of a node in the network depends on the number
of neighbors and the redundancy constant. The usage of a fixed redundancy constant in the network causes unbalanced
transmission load and may cause early depletion of energy sources of nodes with less neighbors. Notice that the number of neighbors is generally
available locally due to the common use of either Neighbor Advertisement/Solicitation control packets or L2 synchronization mechanisms. 
We leverage this fact to introduce multiple redundancy constants among nodes in the network, dependent on the number of neighbors.

Although our first attempt was to derive a closed form expression that will provide a locally optimal value of $K_i$, due to the complexity of
the model this remains an open problem. Instead, we propose a simple calculation of $K_i$ feasible on constrained devices. The idea is to
increment $K_i$ for each redundancy {\it step} number of neighbors. On the other hand, parameter redundancy {\it offset}, specifies the
number of neighbors for each node that corresponds to the minimal value of $K=1$. The calculation is outlined in Algorithm \ref{algo:calculate_k}.

\begin{algorithm}
\caption{Local calculation of the redundancy constant $K_i$}\label{algo:calculate_k}
\begin{algorithmic}[1]
\Procedure{calculate\_k}{$num\_neighbors, step, offset$}
\If {num\_neighbors $\le$ offset}
\Return 1
\Else \State{ \Return $\lceil \frac{num\_neighbors - offset}{step} \rceil$}
\EndIf
\EndProcedure
\end{algorithmic}
\end{algorithm}

\begin{figure*}[tb]
	\centering
	\hfil
	\subfigure[][Offset = 0, Step = 2]{
		\includegraphics[width=0.6\columnwidth]{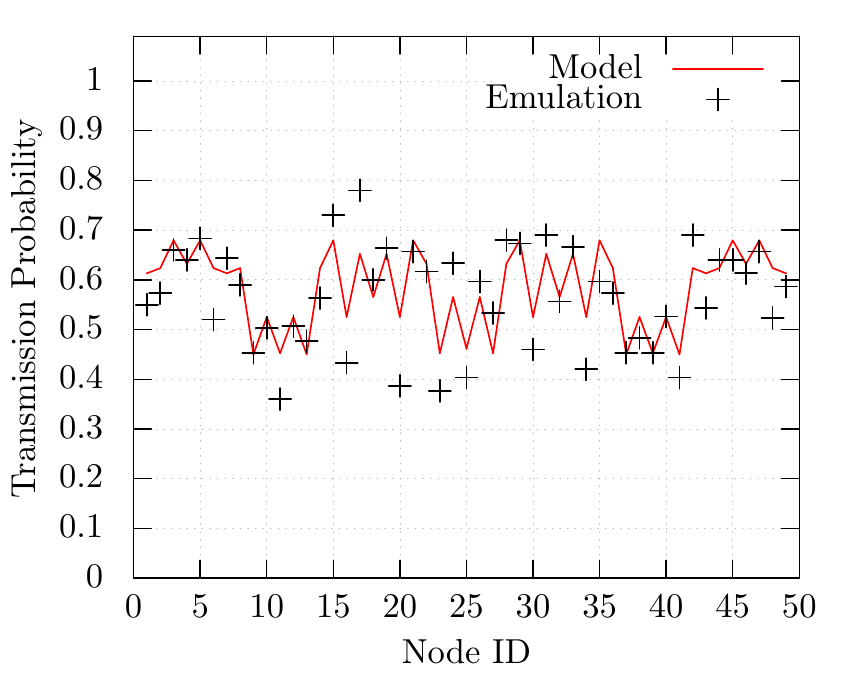}
		\label{per_node_offset_0_step_2}
	}
	\subfigure[][Offset = 0, Step = 3]{
		\includegraphics[width=0.6\columnwidth]{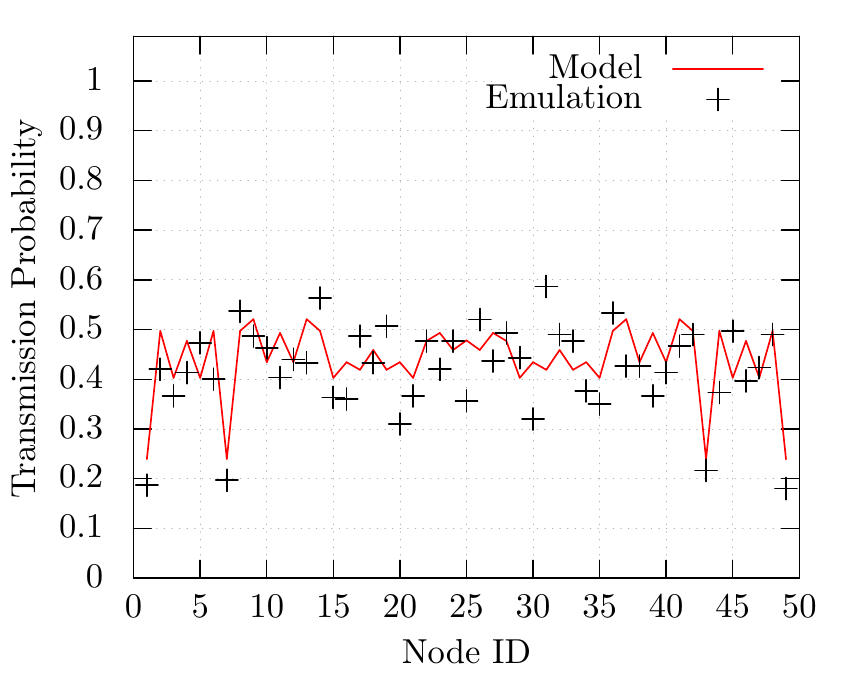}
		\label{per_node_offset_1_step_3}
	}
	\subfigure[][Offset = 2, Step = 3]{
		\includegraphics[width=0.6\columnwidth]{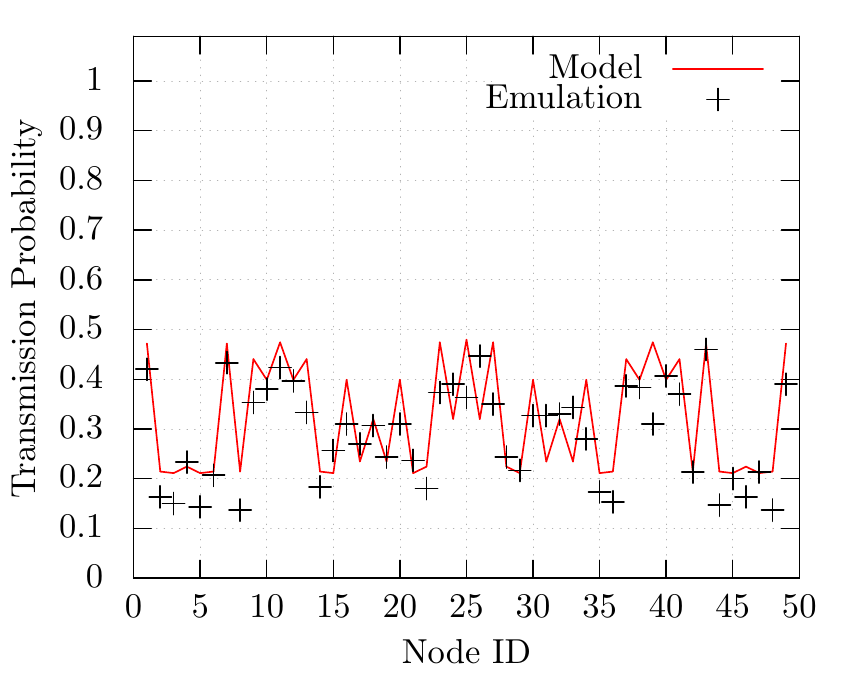}
		\label{per_node_offset_2_step_3}
	}
\caption{Transmission probability of nodes in the grid estimated by the model and compared to emulation results for locally computed redundancy constant.
}
\label{3d_var_k}
\vspace{-0.3cm}
\end{figure*}

We show the effect estimated by the model of the locally computed redundancy constant for the most interesting combinations
of parameters in Fig. \ref{3d_var_k}. We also confront the estimations with emulation results in Table~\ref{var_k_table}.

We can see that the use of the locally computed redundancy constant greatly reduced the effects observed 
in Fig. \ref{3d_const_k}. Depending on the parameters passed to the procedure, we note that the effect can be either
reversed such that the nodes in the corners transmit with smaller probability than the nodes inside the grid, or reduced which is the case
for nodes on the edges of the grid. Clearly, the absolute value of the ideally balanced transmission probability in the network depends on
the requirements of the application actually using Trickle. 
 
 
 \textbf{Parameter Selection.}
 In original Trickle, the redundancy constant $K$ is a parameter that effectively depends on the application requirements.
 With our algorithm, we extend this concept in order to catch the topology characteristics and to provide a better transmission load distribution
 among nodes. However, both redundancy {\it step} and redundancy {\it offset} effectively depend on the application
 using Trickle and are semantically equivalent to $K$. The notion of ''redundancy'' from the application point of
 view is in our case defined as a function of the network topology, i.e. how many transmissions are needed for a given neighborhood to reach
 application needs. 
 For instance, with $step = 2$ and $offset = 0$ application specifies that a transmission should be 
 suppressed when at least half of the neighbors have advertised their state as consistent. 
In parallel, {\it step} regulates the granularity of local $K_i$ increments This directly affects the distribution of the transmission load in 
the network. 
 Thus, instead of blindly defining $K$, 
  the application will have a finer control on the redundancy depending on the topology. In the same time it achieves better a
 transmission load distribution. 
 
 \begin{table}[tb]
 \caption{Model and emulation results for the locally computed redundancy constant on the grid. To obtain $K \in \{1,2\}$, 
we used $offset=2$, $step=3$, and for $K \in \{1,2,3\}$, $offset=0$, $step=3$.} 
 \begin{tabular}{|p{1.8cm}|p{1.2cm}|p{1.2cm}|p{1.2cm}|p{1.2cm}|}
   \hline
   results / redundancy constant & model: $K\in\{1,2\}$  & emulation: $K\in\{1,2\}$  & model: $K\in\{1,2,3\}$  & emulation: $K\in\{1,2,3\}$ \\
   \hline
   average message count & 15.734 & 15.326 & 21.587 & 21.66 \\
   \hline
   max probability & 0.479 & 0.493 & 0.520 & 0.586 \\
   \hline
   min probability & 0.011 & 0.15 & 0.239 & 0.213 \\
   \hline
   Variance & 0.01188 & 0.00947 & 0.00511 & 0.00800 \\
   \hline
 \end{tabular}

 \label{var_k_table}
\vspace{-0.5cm}
\end{table}

\vspace{-0.2cm}
\section{Conclusion}
\label{conclusion}
In this paper we presented a model of the Trickle algorithm that estimates the message count in steady state.
We do this by calculating average transmission probabilities of individual nodes in the network. This allowed 
us to demonstrate load misbalance and unfairness of the algorithm when used with a unique redundancy
constant in the network. The root cause of the unfairness is the heterogeneity of the underlying topology as
nodes do not have the same number of radio neighbors in their range. As a consequence, with a unique redundancy constant,
nodes with less neighbors transmit Trickle broadcast messages more often.
We validated our model by comparing it with emulation results and demonstrated its high accuracy.
In order to improve the fairness of Trickle, we proposed a simple heuristic algorithm that locally computes the redundancy
constant of a node based on the number of neighbors in its vicinity. We demonstrated that by using our algorithm, nodes in the
network achieved better transmission load distribution. However, deriving an optimal value of the redundancy constant
that will perfectly balance the transmission probabilities of nodes in heterogeneous topologies remains an open problem
that we plan to study as future work.

\vspace{-0.1cm}
\bibliographystyle{IEEEtran}
\bibliography{trickle-model}


\end{document}